# An Ontology-based Collaborative Business Intelligence Framework

Muhammad Fahad[1], Jérôme Darmont[1]
[1] *Univ Lyon, Univ Lyon 2, UR ERIC,*
*5 avenue Mendès France, 69676 Bron Cedex, France*
*f.muhammad@univ-lyon2.fr, jerome.darmont@univ-lyon2.fr*

Abstract: Business Intelligence constitutes a set of methodologies and tools aiming at querying, reporting, on-line analytic processing (OLAP), generating alerts, performing business analytics, etc. When in need to perform these tasks collectively by different collaborators, we need a Collaborative Business Intelligence (CBI) platform. CBI plays a significant role in targeting a common goal among various companies, but it requires them to connect, organize and coordinate with each other to share opportunities, respecting their own autonomy and heterogeneity. This paper presents a CBI platform that democratizes data by allowing BI users to easily connect, share and visualize data among collaborators, obtain actionable answers by collaborative analysis, investigate and make collaborative decisions, and also store the analyses along graphical diagrams and charts in a collaborative ontology knowledge base. Our CBI platform builds a dashboard to persist collaborative analysis, supports interactive interface for tracking collaborative session data and also provides customizable features to edit, update and build new ones from existing graphs, diagrams and charts. Our CBI framework supports and assists information sharing, collaborative decision-making and annotation management beyond the boundaries of individuals and enterprises.

## 1 INTRODUCTION

In recent years, there has been a massive and rapid growth of data. The transformation of large volumes of data into useful information to help the decision making process is called Business Intelligence (BI). According to InfoTech research (2020), BI is defined as an enterprise-wide capability to capture, transform and report data or an event into actionable information to enable fact-based tactical and strategic decisions. BI plays a vital role in fulfilling the overwhelming need of large enterprises for analyzing business data in competitive environments. Therefore industrialists and researchers develop BI strategies and tools to enable reporting and analytics for decision making on large datasets, strengthen business processes and operational research activities. BI analytic tools and technologies help reap the maximum benefits from business operations and take good data-driven business decisions. Through well-informed business decisions with real-time data, organizations compete with each other and improve business forecasting in large business clusters.

Typical BI software has features such as reporting and visualization, trend analysis, customer behavior analysis, predictive modeling, etc. However, BI only enables and restricts decision making features within the boundaries of individual companies. In addition, traditional BI tools provide services to individual companies rather than a network of companies characterized by organizational, lexical and semantic heterogeneity. This drawback leads to explore innovative approaches such as Collaborative Business Intelligence (CBI) to make collective decisions incorporating external data beyond the boundaries of enterprises.

There can be many forms of CBI. It may be a general discussion among people within companies or a more report-centric discussion aiming at commenting and providing feedback on a particular report (Tackels, 2015). Other forms of CBI may be seen as adding annotations to specific items in a report, data visualization, or information sharing, etc. In addition, through CBI organizations achieve information sharing on top of BI tools that include data warehouses, analytical tools and reporting tools.

CBI enables organizations to gain timely access to quality information and competitive advantages.

This paper introduces our CBI framework, which is an accessible, open-source BI platform that implements the data warehousing process in Software-as-a-Service (SaaS) mode. CBI users connect to the platform where they can share and explore data, help each other to construct data cubes, and formalize different visualizations via traditional graphs and charts. CBI users can talk with each other, build collaborative analysis, comment and annotate their findings along the compelling graphs and finally store their analysis in the dashboard. We also design a CBI ontology that stores all types of annotations, rather than a database or text files. Ontology-based storage of annotations indeed makes them machine-processable, interpretable and they enable high precision when searching and retrieving knowledge.

The remainder of the paper is organized as follows. Section 2 presents related work on CBI. Section 3 details our CBI platform. Finally, Section 4 concludes this paper and hints at future research.

## 2 RELATED WORKS

There is diverse work in the field of CBI. We classify the existing literature on CBI into several categories. The first category consists of research works focused on collaborative query management. Giacometti et al. (2011) recommend multidimensional queries based on a distance measure calculated by comparing log sessions and the current session. However, they do not consider user context or preferences, which are addressed by Eirinaki et al. (2014). Their system is based on a lookup mechanism where similar users and queries help recommend queries to the current user. Khoussainova et al. (2011) propose another research work for auto-completion and query management. They develop a context-aware system that helps novice users formulating SQL queries. Their system does not recommend complete queries, but possible additions to various clauses in the user's queries. Sapia (2000) differs from these approaches, as they use predictive prefetching in on-line analytical processing (OLAP) to minimize query execution time. They use a Markov model based on the user's multidimensional data. Moreover, Golfarelli et al. (2012) enhance the decision making process by sharing knowledge and operational data across networks of peers. They develop a language for semantic mappings between multidimensional schemata of peers, as well as a query reformulation framework.

The second category we propose consists of research works that focus on OLAP session analysis for preference-based recommendations within BI platforms. It may be difficult to differentiate them from the first category, but they actually focus on analyzing previous sessions and not recommend single OLAP queries. Jerbi et al. (2009) investigate the interactive and navigational nature of user query behavior. They extract relevant elements from the user profile and enrich the query answer before presenting it to the end-user. Next, Aligon et al. (2015) present a collaborative filtering approach that recommends OLAP sessions by analyzing previous sessions, while not recommending single OLAP queries. Their work is unique in the sense that they compare sessions rather than queries. Aufaure et al. (2013) also use recent analytical sessions to recommend queries based on a probabilistic user behavior model and query similarity metrics. They aim to reduce latency time by the use of a cache manager that prefetches objects. Wu et al. (2007) combine the aggregation power of OLAP and keyword-driven analytical processing. They develop scalable algorithms for subspace generation, novel ranking and dynamic facet construction. Cabanac et al. (2007) develop relational OLAP to add annotations on multidimensional data. Such annotations enable decision-makers to share and communicate data among all collaborators. This work is very close to ours, but we store annotations in an ontology rather than a database. Ontology-based storage of annotations makes them machine-processable, interpretable and they enable high precision for searching and retrieving knowledge. Ontology-based dashboard let different collaborators to get a better understanding of the data by making it easier for them to find, manipulate and access information related to collaborative sessions held between different collaborators.

## 3 COLLABORATIVE BUSINESS INTELLIGENCE PLATFORM

This section details our CBI platform, which is accessible, open source and free. It implements the data warehousing process in Software-as-a-Service mode. In the following subsections, we first present the architecture of our CBI Framework. Then, we discuss its implementation and elaborate a use case scenario to demonstrate how our CBI framework helps collaborators to work together.

## 3.1 CBI Framework Architecture

The main components of our CBI platform are the collaborative OLAP, Annotation Management System (AMS) and User Session Handler (USH). The AMS and USH components help store collaborative data semantically into the knowledge base. By collaborative data, we mean parameters of data cube (measures, dimensions, filters, etc.), collaborator personal information and annotations (comments, opinion, analysis, etc.) by collaborators. Figure 1 illustrates the architecture of our CBI platform. With these components, our CBI platform democratizes data by allowing BI users to easily connect, share and analyze data, obtain actionable answers and store their collaborative analyses along the graphical diagrams and charts in the collaborative knowledge base.

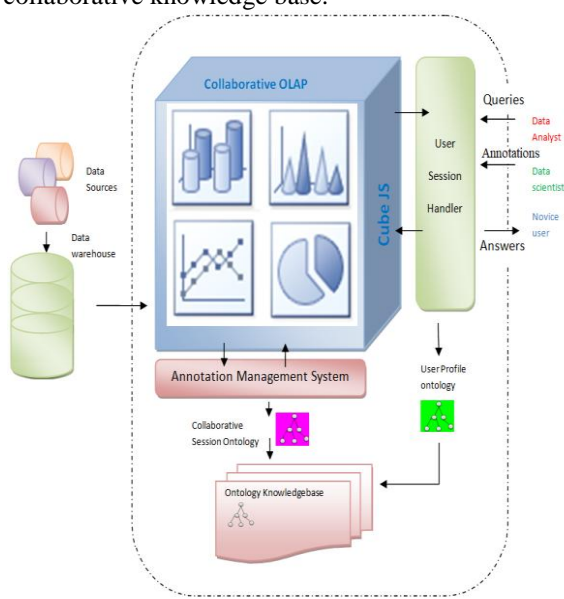

Figure 1. Our CBI platform

### 3.1.1 Collaborative OLAP

The most significant foundation stone in BI is leveraging OLAP, which permits end-users to navigate through aggregated data in a multidimensional data model. It supports various visualization features, such as creating various charts, representing tabular data and also standard operators including drilldown, rollup, and slice and dice. Different end-users, i.e., data analysts, data scientists and novice users, can run OLAP as per their requirements and make a collaborative session to organize, analyze and visualize data multi-dimensionally. First, collaborators can connect to the desired backend database or data warehouse. Then, data cubes schemas can be designed in collaboration with teams according to their requirements. Traditionally, data cubes derive from a relational fact table that contains some quantitative metrics, i.e., measures, over which some calculations can be performed, and dimension tables that are axes of analysis whose attributes are called members. Finally, users can visualize cubes to measure and display dimensions via different types of visualization diagrams. They can *interact* via the USH and *collaborate* over multi-dimensional data via the AMS. Both components are elaborated below.

### 3.1.2 User Session Handler

Collaborators connect to the CBI platform via the USH, where they can collaborate with each other. When a collaborator connects, the USH stores all user-specific information in the User Profile Ontology (UPO). It also stores the location and spatiotemporal information about the collaboration held between collaborators. The UPO allows the reusability of online Web ontologies, i.e., FOAF, (Friend Of A Friend), TimeLine and GeoNames. Particularly, we reuse a FOAF ontology (Vakaj and Martiri,2011) to capture collaborator information in the CBI platform. The FOAF ontology describes persons, their activities and their relations to other people and objects. We also use the TimeLine ontology (Raimond and Abdallah, 2007) that captures the temporal information of collaboration. In addition, we use the GeoNames ontology (Maltese and Farazi, 2013) to capture either the physical or virtual location of the collaborative session. This ontology constitutes a well-known geospatial dataset providing data and metadata from around 7 million unique named places collected from several sources.

### 3.1.3 Annotation Management System

The collaborative session takes input from collaborators over the OLAP graphical interface. Input can be annotations of any type, i.e., question, answer, text comment or description, and can be of any form, i.e., general feedback, report centric discussion, data analysis, task coordination, information sharing, etc. The AMS takes all the annotations related to discussion and analysis held between collaborators, and stores them in the Collaboration Session Ontology (CSO). The AMS allows several operations such as add, edit, update and delete, over collaborator annotations.

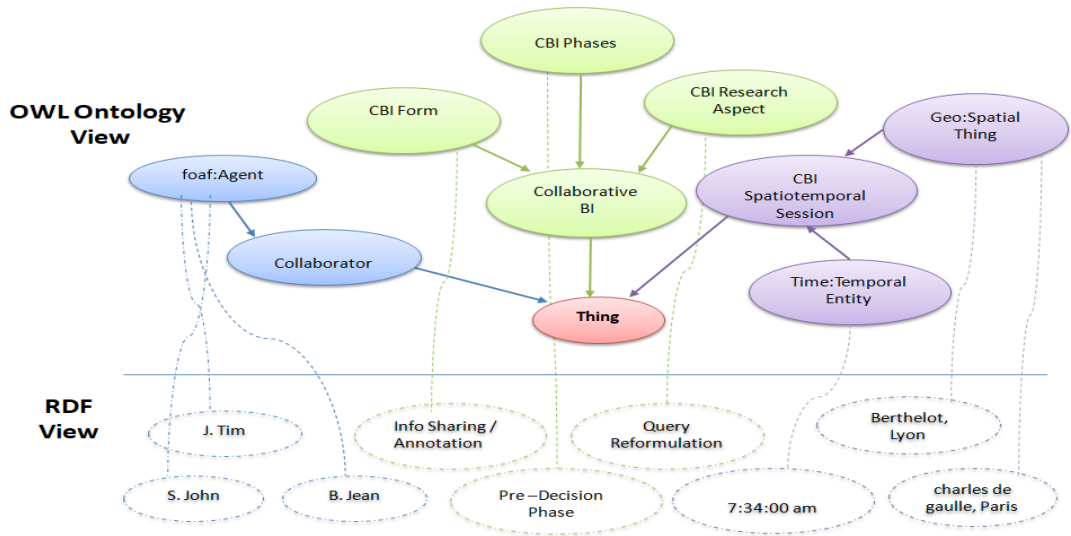

Figure 2. Top level view of CBIOnt

### 3.1.4 CBI Knowledge Base – CBI Ontology

Our aim is to build an efficient and fast storage and retrieval of information between collaborators. Therefore, we aim at incorporating ontologies in the collaborative platform so that different types of inferences can be achieved on collaborative session data. We design the CSO to formally describe and conceptualize the domain knowledge and store collaborative session data between collaborators. Data captured in the CSO becomes machine-interoperable and machine-processable to facilitate easy knowledge sharing, with common vocabulary across independent collaborative teams and organizations. These subontologies UPO and CSO together constitute a CBI ontology named CBIOnt (Fahad et al., 2022).

### 3.2 Implementation of the CBI Framework

We use CubeJS[a] for building our CBI platform. CubeJS is an open source BI platform that supports data integration from all major data sources, designing multidimensional data warehouses and OLAP navigation. Moreover, CubeJS implements the GraphQL (Porcello and Banks, 2018) API that provides a complete and understandable description of the data. CubeJS constructs data cubes that exploit a JSON-based metric skeleton to express data calculations that can be exposed by GraphQL. There are three tabs in the CBI framework through which BI users build their collaborative analysis over OLAP. The details are following.

### 3.2.1 Exploration

The first tab of the interface, "Explore", allows end-users create various types of visual cube representations. It allows selecting measures and displaying dimensions of the data cube. CBI users can apply filters and choose segments and time frames to visualize cube data. This tab also allows users of the collaborative framework to add comments on the cube via the new "Add Comment" button. By the new "Enlist Comment" button, CBI users can see all the comments added during the collaborative session held among collaborators. One can add, edit and delete comments from the collaborative session at any time. Once the cube is formed and comments are added by collaborators, the "Add to Dashboard" button stores the cube on the dashboard so that collaborators can use it later on.

### 3.2.2 Dashboard

The dashboard enables collaborative analysis persistence. The "Dashboard" tab helps store and visualize all the cubes created by collaborators. Moreover, we enhance the dashboard so that already created data cubes can be editable at any time. Each cube is provided with options to edit, delete and enlist comment buttons. On clicking the edit button, the cube enters in the "Explore" interface where updating the cube is possible. When a collaborator saves a cube along with comments, our platform persists his/her analysis for future use.

---

[a] cubejs site: https://cube.dev/

### 3.2.3 Export Data

With the help of the "Export Data" tab, end-user can export all stored cubes on the dashboard as JSON files containing GraphQL queries that allow reconstructing cubes, particular information (name, description, etc.) about cubes, and all the comments added by the collaborators.

## 3.3 Use Case Scenario

We build our case study upon the Star Schema Benchmark (SSB; Neil et al., 2009). SSB provides a data model, i.e., a multidimensional schema (Figure 3) and a workload model, i.e., a set of queries as analyses. Thus, we can build graphs and charts to demonstrate our tool. Let us discuss the SSB aspects that are necessary for understanding our use case scenario.

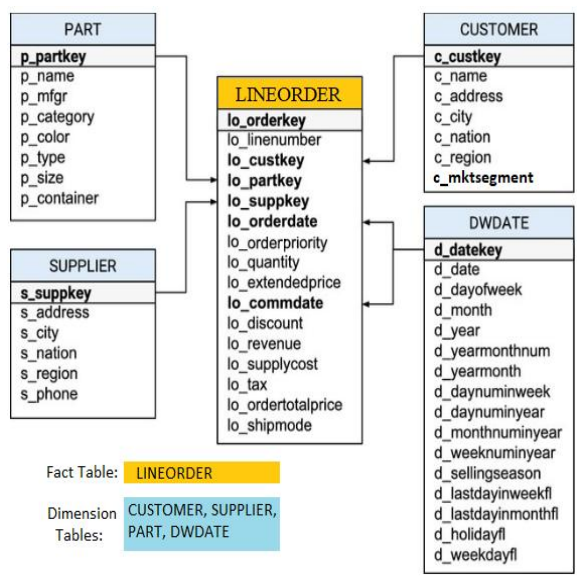

Figure 3. SSB Snowflake Schema

### 3.3.1 Schema and Dataset

In SSB, there are four dimension tables, i.e., CUSTOMER, SUPPLIER, PART and DWDATE, and a fact table named LINEORDER. In data warehousing, a fact table consists of measurements, metrics or facts. The fact table may contain many degenerate dimensions. According to Kimball (2002), in a data warehouse, a degenerate dimension is a dimension key in the fact table that does not have its own dimension table, because all the interesting attributes have been placed in analytic dimensions. Degenerate dimensions are essential for grouping together related fact table's rows. We only focus here on attributes from LINEORDER that are necessary in the upcoming sections.

Measure count calculates the total number of orders.

Degenerate dimension lo_orderpriority is a fixed text. Only five values are allowed: URGENT, HIGH, MEDIUM, NOT SPECIFIED and LOW.

Degenerate dimension lo_shipmode is a fixed text. Only seven values are allowed: AIR, SHIP, MAIL, FOB, TRUCK, RIG AIR, and RAIL.

We use degenerate dimensions lo_shipmode and lo_orderpriority for grouping together related rows in LINEORDER fact table.

### 3.3.2 Use Case

Jean belongs to an organization that casually uses our CBI platform. She meets Kim at the Data Summit and has an exchange together. Kim, who is a novice user, finds interesting to capture knowledge exchanged within collaborative sessions and benefit from data visualization and information sharing. He asks Jean to help him explore some data (actually SSB's data). The conversation during the collaborative session is elaborated below. Kim looks at the LINEORDER fact table and inquires what types of mode of shipment are possible for the delivery of orders, and what types of order priorities are set by customers. While looking at the lo_shipmode attribute in LINEORDER, Jean notices that there are seven types of shipment of orders, i.e., AIR, SHIP, TRUCK, RAIL, etc. Immediately, she creates a cube (Figure 4) that counts the orders and dimension lo_shipmode to display shipment of orders.

Then, Jean chooses a pie chart representation (Figure 5) to observe modes of shipments. She realizes that the most common type of freight transport is TRUCK. Kim and Jean discuss whether road shipping is the most cost-effective way to ship orders. Kim wants to investigate the proportion of shipment modes. Helped by Jean, Kim adds the lo_orderpriority dimension in the cube and chooses a tabular representation of data. Jean tells him that he can create other types of charts (line graph, bar chart, etc.) for better visualization.

```
cube(`Lineorder`, {                         dimensions: {
  sql: `SELECT * FROM                         loLinenumber: {
        ssb.lineorder `                         sql: `lo_linenumber`,
  measures: {                                   type: `string`
    count: {                                  },
      type: `count`,                          loOrderdate: {
      drillMembers:                             sql: `lo_orderdate`,
[loOrderdate, loCommitdate,                     type: `time`
loOrderpriority, loShipmode]                  },
    },                                        loCommitdate: {
    loOrdtotalprice: {                          sql: `lo_commitdate`,
      sql:                                      type: `time`
`${CUBE}.\`lo_ordtotalprice\``,               },
      type: `sum`,                            loOrderpriority: {
      format: `currency`,                       sql: `lo_orderpriority`,
      drillMembers:                             type: `string`
[loOrderdate, loCommitdate]                   },
    },                                        loShipmode: {
    loExtendedprice: {                          sql: `lo_shipmode`,
      sql:                                      type: `string`
`${CUBE}.\`lo_Extendedprice\``,               },
      type: `sum`,                            loShippriority: {
      drillMembers:                             sql: `lo_shippriority`,
[loOrderdate, loCommitdate]                     type: `string`
    },                                        },
    loQuantity: {                             loDiscount: {
      sql:                                      sql: `lo_discount`,
`${CUBE}.\`lo_quantity\``,                      type: `number`
      type: `sum`,                            },
      drillMembers:                           loSupplycost: {
[loOrderdate, loCommitdate]                     sql: `lo_supplycost`,
    },                                          type: `number`
    loRevenue: {                              },
      sql:                                    loTax: {
`${CUBE}.\`lo_Revenue\``,                       sql: `lo_tax`,
      type: `sum`,                              type: `number`
      drillMembers:                           },
[loOrderdate, loCommitdate]                 },
    },  },                                    dataSource: `default`
                                            });
```

Figure 4. Cube for SSB Order

Kim chooses a bar chart and observes order priorities (Figure 5). He observes that Urgent and High demand of deliveries are mostly required when goods need to be shipped right away and must be delivered as fast as possible. Both Jean and Kim add their diagrams onto the dashboard.

Now, Kim is very curious to know whether there is a correlation between mode of shipment and delivery priorities, and what mode of shipment is mostly preferred by suppliers to meet order deadlines? He restricts his interest modes of shipment TRUCK and AIR and delivery types HIGH, LOW and URGENT. Kim comments on the CBI platform. Immediately, Jean updates the cube, creates a bar chart measuring the count of LINEORDER and displays the mode of shipment and order priorities to observe underlying data. Both add filters on order priority and order mode of shipment for the some values they want to investigate (Figure 6).

The conversation goes on between the two friends until Kim clears his opinion about the mode of shipment and delivery priorities of orders. All the conversation between him and Jean happens in the form of comments that are annotated along the graph and stored in the dashboard.

Eventually, Kim wants to share his experience with other team members of his own organization. Jean tells him about the "Export data" feature of the CBI platform, which allows end-users to export a whole dashboard as a JSON file. He exports the dashboard and takes it with him. He can now benefit from the charts, comments and annotations during meetings with his colleagues.

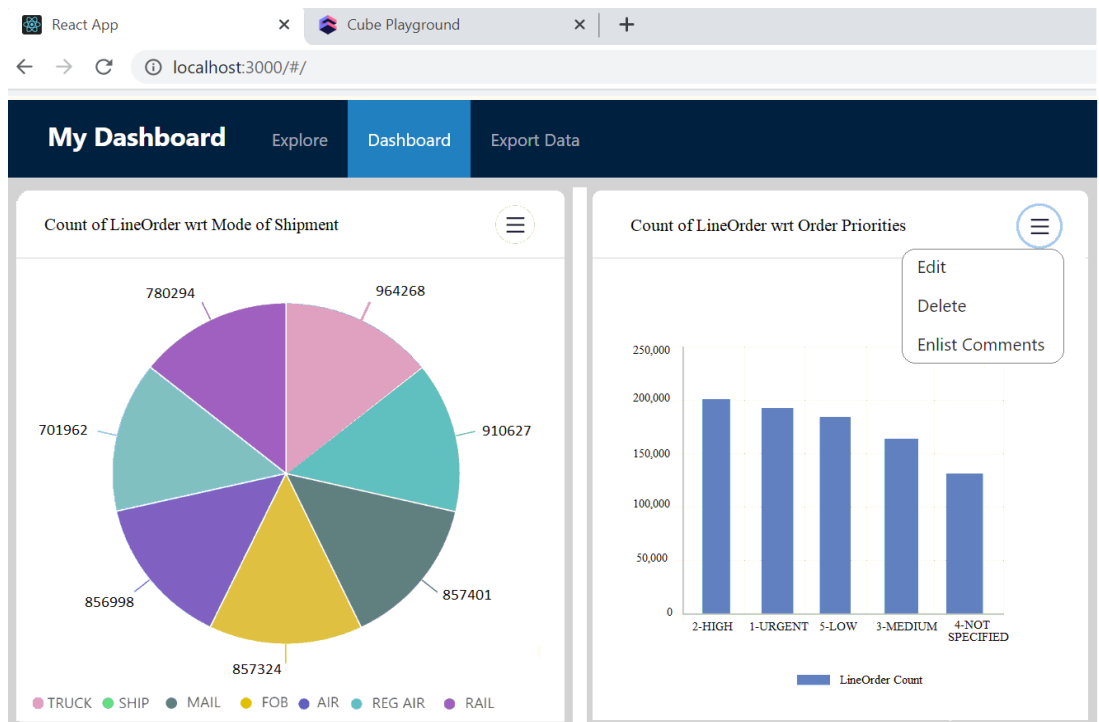

Figure 5. Dashboard containing a pie chart and bar graph

## 4 CONCLUSIONS AND PERSPECTIVES

This paper presents our CBI platform, which enables collaborative data explorations where BI end-users easily connect, manipulate data, uncover hidden facts, make comprehensive overview of data and present their findings in compelling visualizations. CBI platform constitutes a dashboard to persist collaborative analysis, supports interactive interface for tracking collaborative session data and also provides customizable features to edit, update and build new ones from existing diagrams and charts at any moment. With this feature, dashboards are available to other collaborators to quickly and easily see trends and correlations in data anytime and anywhere, consequently achieving time saving benefits. Our CBI platform uses the CBIOnt ontology to store session knowledge on ontologies as open, smart, machine-interoperable and machine-processable data to facilitate easy domain knowledge sharing, with a common vocabulary across independent collaborative teams and organizations. In this way, dashboard connected with CBIOnt is useful for monitoring, measuring and analysing data among collaborators, and also enabling efficient and effective storage and retrieval of session data. One of our ongoing future directions is to make a searchable dashboard based on semantic features. We believe that the semantic layer based on ontologies shall play a major role in CBI's development.

## ACKNOWLEDGEMENTS

The research depicted in this paper is funded by the French National Research Agency (ANR), project ANR-19-CE23-0005 BI4people (Business Intelligence for the people).

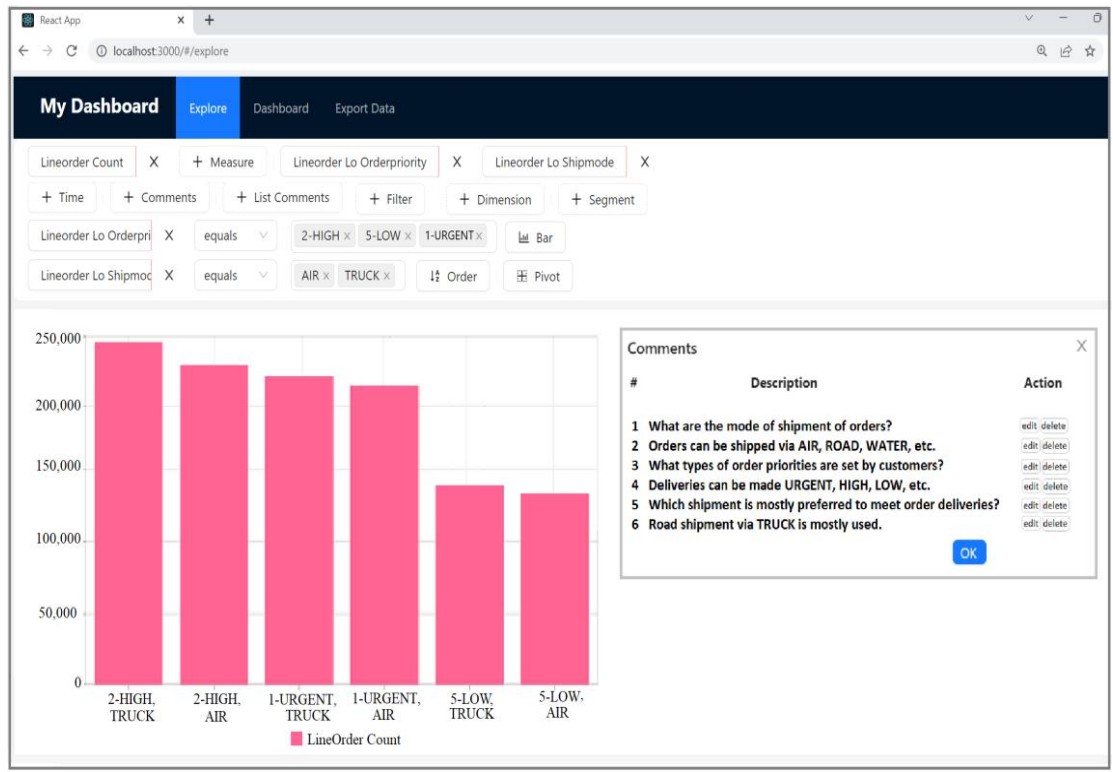

Figure 6. Order count w.r.t. Order Priorities and Order Shipmode

## REFERENCES

Aligon J., Gallinucci E., Golfarelli M., Marcel P., Rizzi S. (2015) A collaborative filtering approach for recom. OLAP sessions, DSS Elsevier vol. 69 pp. 20-30

Aufaure, M.A., Kuchmann-Beauger, N., Marcel, P., Rizzi, S., and Vanrompay, Y. (2013) Predicting Your Next OLAP Query Based on Recent Analytical Sessions, DaWaK, LNCS, vol. 8057, pp. 134–145, Springer

Cabanac, G., Chevalier, M., Ravat, F., Teste, O. (2007) An annotation management system for multidimensional databases. DaWaK, LNCS, pp. 89-98

Eirinaki M., Abraham S., Polyzotis N., Shaikh N. (2014) QueRIE: Collaborative database exploration. IEEE TKDE, vol. 26 (7), pp. 1778-1790

Fahad M., Darmont J., and Favre C (2022) The Collaborative Business Intelligence Ontology (CBIOnt), BI and Big Data, RNTI-B-18, pp 61-72

Giacometti A., Marcel P., Negre E., and Soulet A. (2011) Query recommendations for OLAP discovery driven analysis, IJDWM, IGI Global, vol. 7(2), pp. 66-90

Golfarelli, M., Mandreoli, F., Penzo, W., Rizzi, S., Turricchia, E. (2012) OLAP query reformulation in peer-to-peer DWH. Info. Sys., vol. 37(5), pp. 393-411

InfoTech (2020), Build a Reporting and Analytics Strategy, https://www.infotech.com/research/ss/build-a-reporting-and-analytics-strategy

Jerbi H., Ravat F., Teste O. and Zurfluh G., (2009) Preference-Based Recommendations for OLAP Analysis, DaWaK, LNCS, vol. 5691, pp. 467-478

Kimball, R., Ross, M. (2002). The Data Warehouse Toolkit: The Complete Guide to Dimensional Modeling (Second ed.). ISBN 978-0-471-20024-6.

Khoussainova N., Kwon Y., Balazinska M., and Suciu D. (2011) SnipSuggest: Context-Aware Autocompletion for SQL, VLDB Endowment, vol. 4(1), pp. 22–33

Maltese, V. and F. Farazi (2013). A semantic schema for geonames. In proc of INSPIRE – June 25th 2013.

O'Neil P.E., O'Neil E. J., Chen X., and Revilak S., The Star Schema Benchmark and Augmented Fact Table Indexing, TPCTC'09, LNCS, vol. 5895, pp. 237-252

Porcello E. and Banks A., (2018), Learning GraphQL - Declarative Data Fetching For Modern Web Apps.

Raimond, Y. and Abdallah, S. (2007). The timeline ontology - owl-dl ontology. Technical report, http://motools.sourceforge.net/timeline/timeline.html.

Sapia C. (2000) PROMISE: Predicting Query behavior to Enable Predictive Caching Strategies for OLAP systems, DAWAK 00 UK, pp. 224-233, LNCS

Tackels D. (2015) 6 benefits of a collaborative approach to analytics & bi, https://www.sigmacomputing.com/

Vakaj, E. and E. Martiri (2011). Foaf-academic ontology: A vocab. for the academic community. pp. 440–445.

Wu P., Sismanis Y., Reinwald B. (2007) Towards keyword-driven analytical processing, Proceedings of the ACM SIGMOD, pp. 617–628. ACM